\documentclass[twocolumn,aps,pre,superscriptaddress,showpacs]{revtex4}
\usepackage{amsmath}
\usepackage{graphicx}
\usepackage{multirow}

\begin{document}
\title{Finite Size Effects on Current Correlation Functions}
\author{Shunda Chen}
\affiliation{Department of Physics and Institute of Theoretical Physics
and Astrophysics, Xiamen University, Xiamen 361005, Fujian, China}
\author{Yong Zhang}
\affiliation{Department of Physics and Institute of Theoretical Physics
and Astrophysics, Xiamen University, Xiamen 361005, Fujian, China}
\author{Jiao Wang}
\affiliation{Department of Physics and Institute of Theoretical Physics
and Astrophysics, Xiamen University, Xiamen 361005, Fujian, China}
\author{Hong Zhao}
\email{zhaoh@xmu.edu.cn}
\affiliation{Department of Physics and Institute of Theoretical Physics
and Astrophysics, Xiamen University, Xiamen 361005, Fujian, China}
\affiliation{Collaborative Innovation Center of Chemistry for Energy
Materials, Xiamen University, Xiamen 361005, Fujian, China}

\begin{abstract}
We study why the calculation of current correlation functions (CCFs)
still suffers from finite size effects even when the periodic boundary
condition is taken. Two important one dimensional, momentum conserving
systems are investigated as examples. Intriguingly, it is found that
the state of a system recurs in the sense of microcanonical ensemble
average, and such recurrence may result in oscillations in CCFs.
Meanwhile, we find that the sound mode collisions induce an extra
time decay in a current so that its correlation function decays faster
(slower) in a smaller (larger) system. Based on these two unveiled
mechanisms, a procedure for correctly evaluating the decay rate of
a CCF is proposed, with which our analysis suggests that the global
energy CCF decays as $\sim t^{-\frac{2}{3}}$ in the diatomic hard-core
gas model and in a manner close to $\sim t^{-\frac{1}{2}}$ in the
Fermi-Pasta-Ulam-$\beta$ model.
\end{abstract}

\pacs{05.60.Cd, 44.10.+i, 05.40.-a, 44.05.+e}
\maketitle

{\it Introduction.$-$} In principle, theoretical predictions of statistical
mechanics only apply in the thermodynamic limit. But in practice the
studied systems are always finite and sometimes can be very small, which
is particularly the case in nanoscience. Hence finite size effects should
be carefully analyzed and taken into account. Another important situation
where finite size effects must be considered is to probe thermodynamic
properties of a system by molecular dynamics simulations. As usually the
system size accessible to simulations is small, the simulation results
may significantly deviate.

For a physical quantity $\cal A$, the current correlation function
(CCF) defined as
\begin{equation}
C_{JJ}(t)\equiv\langle J(0)J(t)\rangle
\end{equation}
plays a crucial role for understanding its transport properties. 
Here $\langle\cdot\rangle$ denotes the equilibrium thermodynamic 
average and $J(t)$ is the total current of $\cal A$ at time $t$. The 
time dependence of $C_{JJ}$ reveals how fluctuations of current $J$ 
relax in the equilibrium state and determines the transport coefficient 
of $\cal A$ by the Green-Kubo formula~\cite{1KuboSP}. However, despite 
numerous efforts, general properties of a CCF are still elusive, 
especially in low dimensional, momentum-conserving systems.  For example, 
early hydrodynamic theory predicts that generally a CCF decays in a manner 
of power law, i.e., $C_{JJ}(t)\sim t^{-\gamma}$, with $\gamma$ being half 
of the dimension of the system~\cite{Alder,LongtimeTail}, but studies in 
recent decades suggest that $\gamma$ may depend on some detailed properties 
of a system. In order to clarify this point in one dimensional (1D) case, 
various theoretical methods have been developed~\cite{Lepri,Dharrev,Pros, 
Nara, LepriPRE03, Mai, JSW, BasileHd/2, Pereverzev03, spohn08, Gray, Del, 
Beijeren12} and recent progress~\cite{Del, Beijeren12} suggests that 
$\gamma$ for the heat current correlation function be $\frac{1}{2}$ and 
$\frac{2}{3}$, respectively, for systems with symmetric and asymmetric
inter-particle interactions. As nowadays it is still impracticable to
measure a CCF in laboratories, one has to employ numerical simulations
to check these theories. Unfortunately, existing numerical results do
not agree with each other~\cite{Lepri, Dharrev}, so that a convincing
test is unavailable yet. The main reason for the disagreement  is the
finite size effects induced by the boundary condition inevitably
involved in the simulations.

Therefore, a key task is to overcome the influence of finite size effects.
For this aim the fixed and the free boundary conditions are not favorable,
because strong finite size effects could result from boundary reflections,
manifested as size dependent oscillations in $C_{JJ}$ with a period of
$\frac{2L}{v_s}$, where $L$ and $v_s$ are, respectively, the size and the
sound speed of the system~\cite{Brunet10}. In contrast, the periodic
boundary condition seems to be a better choice. It is free of boundary
reflections and is believed to be effective to suppress finite size
effects~\cite{GangPRB00}. Now the periodic boundary condition has been
extensively adopted in numerical studies~\cite{GangPRB00, wangleiEPL,
LevashovPRL11, Brunet10, Dharcmmt02, Grassberger, Casati03gas, Wang04,
Prosen05} as a convention.  Nevertheless, for momentum conserving systems,
it has been verified that the CCFs numerically obtained with the periodic
boundary condition still have a strong dependence on the system size.
Two general size dependent features are: (i) The CCFs in a smaller
system decay faster than in a larger one~\cite {wangleiEPL, Dharcmmt02,
Grassberger, Casati03gas, Prosen05}, and (ii) size dependent oscillations
with a period of $\frac{L}{v_s}$ instead may occur~\cite{LevashovPRL11,
Brunet10, Dharcmmt02, Grassberger, Casati03gas, Wang04, Prosen05}. The
underlying mechanisms of these phenomena have not been understood yet;
thus how to avoid their effects to achieve trustable numerical results
is still a challenge.

Our aim here is to reveal the mechanisms of these finite size effects
and show how to capture the asymptotic decaying behavior of CCFs
accordingly. We will first show that a local current fluctuation may
excite two pulselike components traveling oppositely and colliding
repeatedly, so that CCFs in a smaller system decays faster due to more
frequent collisions. Then we will show that a finite, momentum conserving
system, has a novel recurrence property even in the equilibrium state.
This recurrence may have more fundamental implications, but for our aim
here we will show that it induces the size dependent oscillations in
CCFs when the current has correlation with the system configuration.
Based on these studies, we will finally propose a reliable procedure
to obtain CCFs free from these finite size effects.

{\it Models and features of CCFs.$-$} We consider two paradigmatic
1D models as illustrating examples: the diatomic hard-core gas model
\cite{Casati86} and the Fermi-Pasta-Ulam-$\beta$ (FPU-$\beta$) model
\cite{FPU}, representing 1D fluids and lattices, respectively. We focus
on the CCF of the total energy current in this work, though our analyzing
method can be applied equally to other currents and quantities. The gas
model consists of $N$ hard-core point particles arranged in order with
alternative mass $m_{o}$ for odd-numbered and $m_{e}$ for even-numbered
particles. The particles travel freely except elastic collisions with
their nearest neighbors. Without loss of generality, we follow Ref.
\cite{Grassberger} to adopt $m_{o}=1$, $m_{e}=3$, and the total energy
current definition $J\equiv \sum j_{i}$, where $j_i\equiv \frac{1}{2}
m_i v_{i}^{3}$ with $m_i$ ($v_i$) being the mass (velocity) of the $i$th
particle. The FPU-$\beta$ model consists of $N$ point particles as well,
defined by the Hamiltonian
\begin{equation}
H=\sum_{i}\frac{p_{i}^2}{2m_i}+V(x_{i}-x_{i-1})
\end{equation}
with $V(x)=\frac{x^{2}}{2}+\frac{x^{4}}{4}$, where $x_{i}$ is the
displacement of the $i$th particle from its equilibrium position and
$p_{i}$ is its momentum. In this model, all particles are assumed to
have a unit mass; i.e., $m_i=1$, and the energy current is defined as
$J\equiv \sum j_i$ with $j_i\equiv \frac{p_i}{m_i}\frac{\partial}
{\partial x_i} V(x_{i+1}-x_{i})$~\cite{Dhar}. In both models the periodic
boundary condition is imposed. The system size $L$ is set to be $L=N$
so that the density of the particle number is unity. In our simulations
the initial condition is set randomly but with two restrictions: The average
energy per particle is unity and the total momentum of the system is
zero. It should be noticed that due to the null total momentum, the
energy current is identical to the heat current~\cite{Lepri}; hence
our discussions on the energy current in the following applies
without any distinctions to the heat current.

In Fig.~1  we plot the energy CCF of the two models. The common
finite size effect shared by them is that the smaller the system
is, the faster the energy CCF decays. Another finite size effect
is observed only in the gas model, which appears as oscillations
whose period depends on the system size.

\begin{figure}
\hskip-.5cm \includegraphics[scale=0.78]{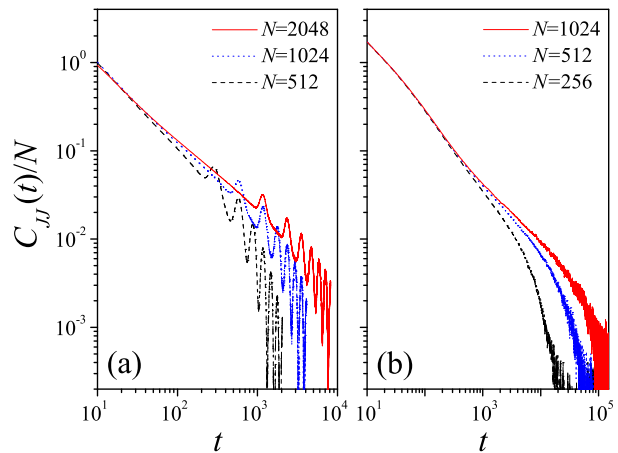}\vskip-0.5cm
\caption{(Color online) The energy current correlation function for the
gas model (a) and the FPU-$\beta$ model (b). The period of oscillations
observed in (a) is measured to be $\frac{L}{v_s}$ ($v_s$ is the sound
speed of the system). }
\end{figure}

{\it Sound mode collisions.$-$} In order to reveal the mechanisms
of these finite size effects, let us first study the spatiotemporal
correlation function of local energy currents. We divide the system
into $\frac{L}{b}$ bins in space of equal width $b=1$. The local
energy current in the $k$th bin and at time $t$ is defined as
$J^{\rm loc}(x,t)\equiv \sum_i j_i (t)$, where $x\equiv kb$ and the
summation is taken over all particles reside in the $k$th bin at time
$t$. The spatiotemporal correlation function of local currents is
defined as~\cite{LevashovPRL11, Casati03gas, Prosen05, Lepri98,
Zhao06, Diffusion13}
\begin{equation}
C(x,t)\equiv\langle J^{\rm loc}(0,0)J^{\rm loc}(x,t)\rangle.
\end{equation}
Note that $J(t)={\sum}_k J^{\rm loc}(kb, t)$, hence we have
$C_{JJ}(t)=\frac{L}{b^2}\int C(x,t)dx$ considering that our systems
are homogeneous in space \cite{Prosen05}. It has been found that
$C(x,t)$ features a pair of pulses moving oppositely away from $x=0$
at the sound speed \cite{Prosen05, Lepri98, Diffusion13}, which are
recognized to be the hydrodynamic mode of sound.

\begin{figure*}[!]
\vskip0.2cm \includegraphics[scale=0.85]{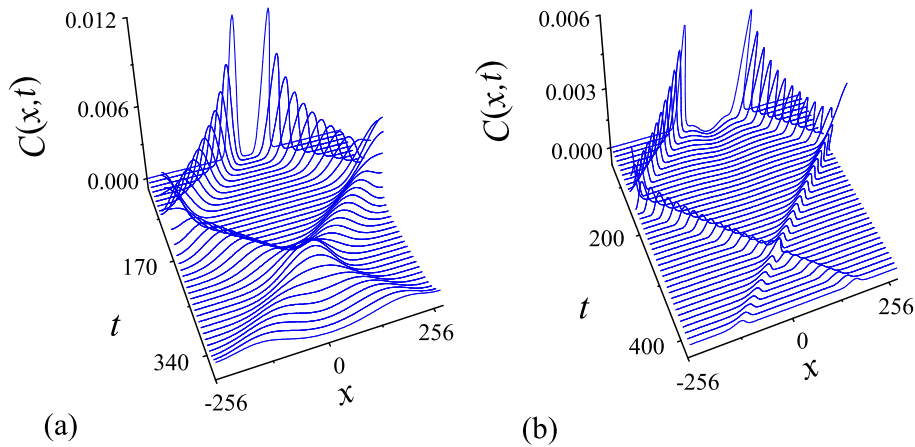}
\vskip-.4cm
\caption{(Color online) The spatiotemporal correlation function
of local energy (heat) currents for the 1D gas model (a) and the
FPU-$\beta$ model (b). The system size is $L=512$ in both cases.}
\end{figure*}

In Fig.~2 we show $C(x,t)$ at various times for both models. The two peaks
representing the sound mode can be clearly identified. Their moving speed
is measured to be $v=1.75$ in the gas model and $v=1.50$ in the FPU-$\beta$
model, agreeing with the sound speed in each system very well. $C(x,t)$
provides more useful information of local currents. As suggested by Fig.~2,
a local current, say, at $x=0$, will excite local currents centered at
$x=\pm vt$ after time $t$. As $C(x,t)$ is positive around $x=\pm vt$, the
excited local currents also have the same flowing direction of the original
local current. If the system has an infinite size, these excited local
currents will never encounter, but in a finite system they will collide
with each other repeatedly (see Fig.~2). Unless they do not interact,
their collisions will damage the excited local currents and in turn cause
the total current $J$ to decay. The collisions take place more frequently
in shorter systems, implying that CCFs should decay faster, in good
consistence with the result presented in Fig.~1. However, the sound mode
collisions can not explain the size dependent oscillations of $C_{JJ}(t)$
observed in the gas model, because they take place with a period of
$\tau_{\text{col}}\equiv\frac{L}{2v_s}$, while $C_{JJ}(t)$ oscillates
with the period of $\frac{L}{v_s}$ [see Fig. 1(a)].


\begin{figure}[b!]
\hskip-.5cm \includegraphics[scale=0.78]{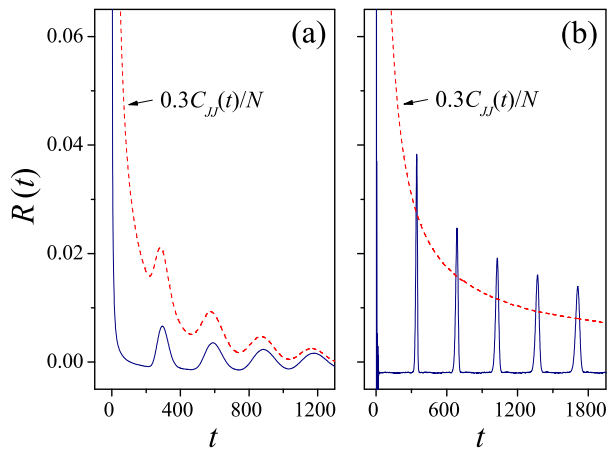}\vskip-.5cm
\caption{(Color online) The comparison between the state correlation
function $R(t)$ (blue solid line) and the energy current correlation
function $C_{JJ}(t)$ (red dashed line) for the gas mode (a) and the
FPU-$\beta$ model (b). In order to have a close comparison, $C_{JJ}(t)$
is shifted down by a factor of $\frac{0.3}{N}$. The system size adopted
is $L=512$ in both pannels.}
\end{figure}

{\it Recurrence.$-$} We find that there is an interesting recurrence
in both systems and it is responsible for the oscillations of $C_{JJ}(t)$
in the gas model. Let us denote the state of a system at time $t$ by
the vector $\mathbf{r}(t)\equiv[p_{1}(t),..., p_{N}(t), x_{1}(t), ...,
x_{N}(t)]^{T}$ and study the state correlation function
\begin{equation}
R(t)\equiv \langle\mathbf{r}(0)\cdot\mathbf{r}(t)\rangle
\end{equation}
to explore the recurrence. The results are shown in Fig.~3: in the sense
of ensemble average, both systems recur with a period of $\tau_{\text{rec}}
\equiv \frac{L}{v_s}$. In addition, the features of $R(t)$ in the gas model
match very well with those of $C_{JJ}(t)$ [see Fig.~3(a) for a comparison],
suggesting strongly that the oscillations of $C_{JJ}(t)$ is closely related
to the recurrence. Indeed, in the gas model $\langle j_{i}(t)x_{i}(t)\rangle
=\langle\frac{1}{2}m_{i}v_{i}^{3}x_{i}\rangle =0$ but $\langle j_{i}(t)p_{i}
(t)\rangle = \langle \frac{1}{2} m_{i}^{2}v_{i}^{4}\rangle=2e$, where $e$
is the average energy of a particle that has been set to be unity, showing
that local currents and the system's momentum configuration are definitively
correlated. However, this is not the case in the FPU-$\beta$ model, where
it has been numerically checked and verified that not only $\langle j_{i}
(t)x_{i}(t)\rangle =0$, but also $\langle j_i(t)p_i(t) \rangle=0$. This
explains why the recurrence does not cause oscillations of $C_{JJ}(t)$
in this model.

{\it Evaluating decaying rates of CCFs.$-$}
Based on the analysis of the two finite size effects, it can be concluded
that how a CCF decays in the thermodynamic limit can only be evaluated by
extrapolating $C_{JJ}(t)$ in the time range of $t\leq\tau_{\text{col}}=
\frac{L}{2v_s}$. Beyond this range, even though the extra decay induced
by each sound mode collision may be weak, the cumulative effect of multiple
collisions can be significantly large. For $t\leq\tau_{\text{col}}$, if the
current has no correlation to the system configuration as in the FPU-$\beta$
model, the function $C_{JJ}(t)$ calculated with a finite system size $L$
should agree with the true CCF in the thermodynamic limit. Otherwise,
as in the case of the gas model, the finite size effect caused by the
recurrence may manifest. But in general, the deviation of $C_{JJ}(t)$
evaluated with a finite system size will decrease as the the system
size increases.

\begin{table}[t!]
\begin{tabular}{c|ccccc}
\hline\hline
\multirow{2}{*}{}
& \multicolumn{5}{c}{system size $L$}\\
\cline{2-6}
model & $~4096~~$ & $~8192~$ & $~16384~$ & $~32768~$ & $~65536$\\ \hline
gas $(t_{0}=100)$ & $0.725$ & $0.699$ & $0.683$ & $0.679$ & $0.672$\\ 
gas $(t_{0}=1000)$ & $0.762$ & $0.704$ & $0.691$ & $0.681$ & $0.675$\\ 
~FPU-$\beta$ $(t_{0}=1000)~$ & $0.614$ & $0.608$ & $0.588$ & $0.576$ & 0.525\\
~FPU-$\beta$ $(t_{0}=2000)~$ & --- & $0.548$ & $0.577$ & $0.568$ & 0.517\\
\hline\hline
\end{tabular}
\caption{The decaying rate $\tilde \gamma$ of the energy CCF for the
two systems measured by the best fitting to $C_{JJ}(t)$ over the time
interval of $(t_{0},\tau_{\text{col}})$. The relative error of $\tilde
\gamma$ ranges from $0.3\%$ to $0.6\%$.}
\end{table}

On the other hand, in order to facilitate the evaluation of the decay
rate, a helpful but not necessary `trick' is to exclude the initial
transient stage of $C_{JJ}(t)$ in performing the extrapolation. As there
is no theoretical guide yet, the time the transient stage lasts, denoted
by $t_0$, has to be determined in a practical way. So we suggest the
following procedure to obtain the decaying rate, $\gamma$, of a CCF: Set
a tentative transient time $t_0$ and measure the decaying rate $\tilde
\gamma$ as a function of both $t_0$ and $L$ by best fitting $C_{JJ}(t)$
over the time range of $(t_{0},\tau_{\text{col}})$, then study the dependence
of $\tilde \gamma$ on $t_0$ and $L$, and identify $\gamma$ to be the value
of $\tilde \gamma$ invariant of further increasing of $t_0$ and $L$ after
certain values. Tab. 1 shows the value of $\tilde \gamma$ for the energy
CCF in both systems; one can find that as expected, the transient time may
affect the convergence rate of $\tilde \gamma$ but does not affect the value 
it tends to. The data presented in Tab. 1 give a strong support that the 
energy CCF decays as $C_{JJ}(t)\sim t^{-\gamma}$ with $\gamma=\frac{2}{3}$ 
for the gas model. For the FPU-$\beta$ model, though the decay rate tends 
to $\frac{1}{2}$ from above monotonously, its value has not converged even 
when the system size is as large as $L=65536$. However, it can be anticipated 
that $\gamma$ for the FPU-$\beta$ model is very likely different from 
$\frac{3}{5}$ as having been constantly concluded in previous numerical 
studies~\cite{Lepri, wangleiEPL, Lepri98}. As a comparison, in Fig.~4(b) 
we plot $t^{\gamma^*}C_{JJ}(t)$ with $\gamma^*=\frac{1}{2}$ and  
$\frac{3}{5}$, respectively; their tendency difference can be 
clearly distinguished.

The study of Tab.~1 suggests that only in a very large system (to the
simulations) may the stationary asymptotic decaying rate of $C_{JJ}(t)$ be
revealed. For example, in order to reach the precision to reliably
distinguish which value $\gamma$ may take among $\frac{1}{2}$,
$\frac{3}{5}$, and $\frac{2}{3}$ in the FPU-$\beta$ model, the system
size should be at least about $L=65536$ which has never been attempted
by previous studies. Moreover, in previous studies, without a careful 
analysis of the finite size effects, the time range for evaluating
$\gamma$ had to be chosen empirically. This may result in big uncertainty
in the results and make it hard to avoid the bias due to existing
theoretical predictions or subjective factors.

{\it Summary and Discussions.$-$} To summarize, we have shown that in
the equilibrium state, a momentum conserving system may recur in the sense
of the ensemble average~\cite{notePhi4} after a size dependent period of
$\tau_{\text{rec}}=\frac{L}{v_s}$. This recurrence may induce oscillations
of a CCF if the current is correlated to the system configuration. It also
implies long time correlations in the system. We emphasize that unlike the
well known Fermi-Pasta-Ulam recurrence~\cite{FPU, ChaosFPU} and the
Poincar{\'e} recurrence~\cite{PoincareRec}, the recurrence found here
features the finite size of a system, the equilibrium state and the
ensemble average,  hence is distinct from them in nature (e.g., its
characteristic recurrence time $\tau_{\text{rec}}$ is much shorter than
that of the FPU and the Poincar{\'e} recurrence). Besides the energy CCF,
this recurrence should have effects on other quantities as well, as long
as they are correlated to the system configuration. For this reason, it
would be interesting to figure out the role it plays in any other related
statistical and dynamical properties.

\begin{figure}[!t]
\vskip-.5cm \includegraphics[scale=0.8]{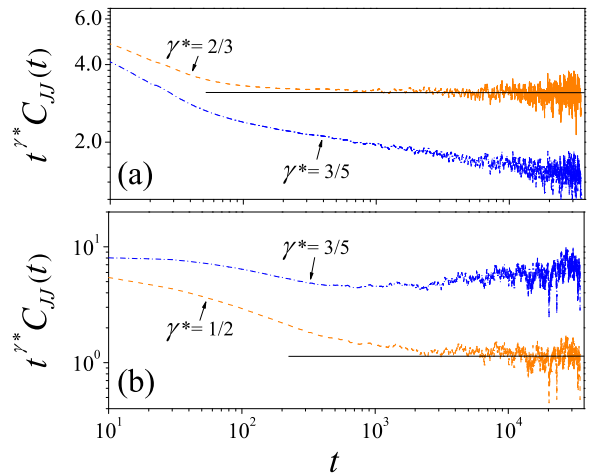}\vskip-.5cm
\caption{(a) The comparison of the rescaled energy CCF, $t^{\gamma^{*}}
C_{JJ}(t)$, for the gas model with $\gamma^{*}=\frac{2}{3}$ (orange
dashed line) and $\gamma^{*}=\frac{3}{5}$ (blue dash-dotted line).
The black horizontal line is plotted for reference. The system size
is $L=65536$. (b) The same as (a) but for the FPU-$\beta$ model with
$\gamma^{*}= \frac{3}{5}$ and $\gamma^{*}=\frac{1}{2}$ instead. The
system size is $L=131072$.}
\end{figure}

In addition, we have shown that the sound mode collisions occurring with
a period of $\tau_{\text{col}}=\frac{L}{2v_s}$ can induce extra scattering
to currents and this explains why a CCF in a smaller system decays faster.
By taking into consideration of these findings, we have proposed a procedure
for reliably measuring the asymptotic decaying rate of a CCF. Our analysis
has suggested that the energy CCF decays as $\sim t^{-\frac{2}{3}}$ in the
gas model and in a manner close to $\sim t^{-\frac{1}{2}}$ in the FPU-$\beta$
model, in agreement with recent theoretical predictions that, in 1D momentum
conserving systems, the energy CCF should decay as $\sim t^{-\frac{2}{3}}$
and $\sim t^{-\frac{1}{2}}$ when the interparticle interactions are
asymmetric and symmetric, respectively~\cite{Del,Beijeren12}. [Note
that the gas (the FPU-$\beta$) model belongs to the class of asymmetric
(symmetric) interactions.]

Nevertheless, we recall that in 1D momentum conserving {\it lattices}
with asymmetric interactions, the heat conductivity may converge and the
heat CCF may decay more rapidly~\cite{Zhong, Chen,ChenASY}. This fact implies that 
there may be some crucial difference between fluid and lattice systems, 
and the decaying prediction of $\sim t^{-\frac{2}{3}}$ may only apply 
to fluids. This point should be clarified in future.

{\it Acknowledgements.$-$} This work is supported by the NSFC (Grants
No. 109255256, No. 11275159, and No. 10805036) and SRFDP (Grant No.
20100121110021) of China.


\begin{thebibliography}{10}
\bibitem{notePhi4}We have checked and found that in momentum nonconserving
systems, such as the $\phi^{4}$ lattice, such recurrence does not occur.
\end{thebibliography}
\end{document}